\documentclass[sigconf,authorversion,balance=false]{acmart}
\usepackage{confstyle}

\copyrightyear{2024}
\acmYear{2024}
\acmISBN{}
\acmConference{arXiv preprint}
\usepackage[T1]{fontenc}
\usepackage{graphicx}
\usepackage{appendix}

\usepackage{multirow}
\usepackage{listings}
\usepackage{xurl}

\usepackage{array}
\newcolumntype{C}{>{\centering\arraybackslash}m{1.5cm}}

\usepackage{pifont}
\newcommand{\no}{\textcolor{red}{\ding{53}}}
\newcommand{\yes}{\textcolor{green}{\ding{51}}}

\begin{document}

\date{}
\title[Stop Following Me!]{Stop Following Me! Evaluating the Effectiveness of Anti-Stalking Features of Personal Item Tracking Devices}

\author{Kieron Ivy Turk}
\email{kieron.turk@cl.cam.ac.uk}
\orcid{0000-0002-4705-4749}
\author{Alice Hutchings}
\email{ alice.hutchings@cl.cam.ac.uk}
\orcid{0000-0003-3037-2684}
\affiliation{%
  \institution{University Of Cambridge}
  \streetaddress{15 JJ Thompson Avenue}
  \city{Cambridge}
  \state{Cambridgeshire}
  \country{United Kingdom}
  \postcode{CB3 0FD}
}
\renewcommand{\shortauthors}{Turk and Hutchings}

\begin{abstract} 
Personal item tracking devices are popular for locating lost items such as keys, wallets, and suitcases. Originally created to help users find personal items quickly, these devices are now being abused by stalkers and domestic abusers to track their victims' location over time. Some device manufacturers created `anti-stalking features' in response, and later improved on them after criticism that they were insufficient. We analyse the effectiveness of the anti-stalking features with five brands of tracking devices through a gamified naturalistic quasi-experiment in collaboration with the Assassins' Guild student society. Despite participants knowing they might be tracked, and being incentivised to detect and remove the tracker, the anti-stalking features were not useful and were rarely used. We also identify additional issues with feature availability, usability, and effectiveness. These failures combined imply a need to greatly improve the presence of anti-stalking features to prevent trackers being abused.

\end{abstract}


\maketitle

\section{Introduction}
Personal item tracking devices such as the Apple AirTag and Tile tracker have recently grown in popularity. These coin-sized devices are intended to be attached to belongings such as keys and bags to allow the owner to quickly find them when they are nearby, or to allow them to be remotely tracked if lost or stolen. Many different brands have started creating their own item tracking devices following the success of the Apple AirTag.

While these devices are useful for their intended purpose, they have begun to be used maliciously. News outlets have reported many cases of AirTags and similar devices being used for stalking and domestic abuse~\cite{motherboard22police}. In late 2022, Apple were sued for this abuse of their devices~\cite{nyt22applesued}. Other malicious uses of item tracking devices have also been reported, such as planting trackers on cars to track where they are parked later to steal them~\cite{macrumors21cartheft,nyt21cartheft}.

Manufacturers are aware of these issues and have implemented a range of ``anti-stalking features'' in response. These commonly include scanning for unwanted trackers following a user and alerting the user of the unwanted tracker. Most implementations additionally provide means to locate the tracker after alerting the user, such as using Bluetooth to indicate the distance to the tracker or making it play a sound alert.

These features later came under criticism for not being sufficient to detect unwanted trackers. Common issues included the sound alerts being too quiet, separated-from-owner alerts taking too long to trigger, and only providing anti-stalking features for certain platforms. Apple later improved on their anti-stalking features following this feedback~\cite{apple22improvements}, however, many believe there is still room for improvement.

We set out to evaluate the effectiveness of the anti-stalking features of tracking devices. We wanted our evaluation to be as naturalistic as possible, but there are also important ethical and data protection issues to consider. We wanted our participants to consent to being tracked, but not necessarily know if or when they were being tracked, by whom, or with what type of tracker. 

We designed our evaluation around a long-established student society, the Assassins' Guild. The Guild runs a competition that spans the entire term. Students enroll in the game and are given a list of targets (other players) to `assassinate'. They also become targets but do not know who their adversary is. The Guild agreed to run a sub-game of participants who agreed to be tracked as part of this evaluation. 

In this paper, we provide an overview of the background and related work in \S\ref{background}. We then discuss the trackers chosen for this study, alongside their features for regular use and stalking prevention in \S\ref{sec:trackers}. Our research was approved by the department's Research Ethics Committee. We provide a detailed overview of the ethical considerations that were central to our research design in \S\ref{ethics}. We provide an overview of our methods in \S\ref{methods}. Our findings are detailed in \S\ref{results}. In our discussion (\S\ref{discussion}) we consider the implications of our study before concluding (\S\ref{conclusion}).

We make a novel methodological contribution for evaluating the use and effectiveness of anti-stalking features. We find that available anti-stalking features do not appear sufficient to prevent many cases of unwanted stalking by tracking devices. We make a number of recommendations so that the features provided by all manufacturers can be improved upon in the future.

\section{Background and Related Work}
\label{background}

\subsection{Technology-Enabled Domestic Abuse}
Technology has become commonplace in domestic abuse, now called technology-facilitated abuse or ``tech-abuse''. Refuge~\cite{refuge19techabuse} found that 72\% of abuse cases involved tech-abuse, while The Safety Net Project~\cite{safetynet15glimpsefromthefield} found 97\% of domestic violence victims reporting technology being misused for abuse.

Multiple studies have explored how technology is used within abusive relationships. Matthews et al.~\cite{matthews17storiesfromsurvivors} provide a framework for analysing technology in domestic abuse. They identify three phases of abusive relationships: physical control, where abusers have full access to the victim's systems; escape, where the victim must hide their actions while severing ties with the abuser; and life apart, where the survivor builds and maintains a new life while avoiding returning to the physical control phase.

Freed et al.~\cite{stalkers-paradise} interview survivors and domestic abuse professionals to identify four categories of tech-abuse: ownership-based access, account or device compromise, harmful messages and posts, and exposure of private information. They describe abusers as the ``UI-bound adversary'' who only use the provided features of a device. This differs from the prior idea of technically skilled abusers. Douglas et al.~\cite{douglas19womensexperiences} interviewed 55 survivors of coercive control and found abusers using technology to isolate victims from their support networks, harass them online, monitor their activity, and stalk them. Dragiewicz et al.~\cite{dragiewicz19dvandcomms} find similar types of technology-facilitated abuse, adding that the use of technology for abuse either starts or escalates in the escape phase of abuse~\cite{matthews17storiesfromsurvivors}.

Leitão~\cite{leitao21techabuseforums} analysed posts on online forums to explore how technology is used for domestic abuse. They find that the key issues are monitoring and stalking in addition to harassment and intimidation as identified by other studies. Lopez-Neira et al.~\cite{lopezneira19abusegettingsmarter} discuss how the problem extends to IoT devices instead of being constrained to mobile phones and personal computers. Finally, Sugiura et al.~\cite{sugiura2021cmaabuse} describe different misuses of technology for abuse, in addition to a list of the monitoring and tracking technologies that they found domestic abusers using. They additionally explore the motivations and justifications by abusers, existing support for victims, and ``hidden groups'' who are often overlooked because they do not fit the typical profile of victims and abusers.

\subsection{Technology-Enabled Surveillance}
A persistent issue is Intimate Partner Surveillance (IPS). Sambasivan~\cite{sambasivan19dontleaveusalone} found cyberstalking to be the main issue in South Asia, with 66\% of their participants reporting instances of cyberstalking. Only 15\% reported impersonation attacks and 14\% reported personal information leaks. Woodlock~\cite{woodlock17techdvandstalking} find technology is used to impose a sense of the abuser's omnipresence, and that the extent of the stalking forced the victims to isolate themselves to be able to escape. Furthermore, the abuser's surveillance provided them with material to punish and humiliate the victims by sharing private information online.

Abusers often make use of spyware and ``dual-use apps'' which provide a legitimate use in addition to an illegitimate one. Chatterjee et al.~\cite{chatterjee18spywareipv} find that the majority of the problem is dual-use apps. Furthermore, these applications often condone their use for IPS, even advertising it as a possible use case. Obada-Obieh et al~\cite{obadaobieh22sokdualnaturesa} focus on dual-use applications in the context of sexual abuse, and highlight how technology can lead to re-victimisation by enabling offenders.

Levy and Schneier~\cite{levy20privacyrelationships} look at broader ``relationships'' where privacy is affected by misuses of technology. This includes IPS in addition to parent-child relationships, adults and the elderly, caregivers, and friends. They identify several common features of the threats posed and relate these to a series of implications for the design of future systems, which aids systems designers who wish to mitigate some of these privacy problems. This includes recognising that there is often a balance between multiple interests and values. This is particularly relevant for location trackers, where tracking may be justified by users due to safety concerns. 

Bellini et al.~\cite{bellini21privacybreedsevil} explore the justifications for IPS on online infidelity forums. They find that prior instances or suspicions of cheating are common justifications, as well as ``suspicious behaviour'' by the victim. They present a four stage cycle of IPS, where abusers often decide on IPS as a ``necessary hardship''. Posters often share their stories to find advice and guidance on how to perform IPS, as well as to share or boast about their surveillance.

\subsection{Usability of Privacy Features}
Privacy features such as the anti-stalking features discussed in this study have important usability choices in their design. These can be designed for the users' benefit, making it easier to improve their own privacy, or to their detriment but to organisations' benefit, allowing for easier data collection. Features which are created for users' benefit should be designed in a manner which guides users towards using them.

Frik et al.~\cite{frik22privacysettings} explore user's expectations of privacy features on their phones. They find that most users have not configured the settings and are unaware of the defaults, although they therefore rely on the defaults in place. Furthermore, users are not aware of which settings are available, and there are some demographic impacts on users' awareness of the features such as older participants interacting with privacy settings less often.

A common feature which aims to aid user privacy are nudges: designing decisions which ``nudge'' users towards certain choices, without removing any options. Caraban et al.~\cite{caraban19nudges} perform a systematic review of literature on nudge designs, identifying 23 mechanisms to nudge users towards better decisions. Acquisti et al.~\cite{acquisti17nudges} discuss how these nudge designs can be used to aid users towards beneficial decisions. This includes providing more information to users to base their decisions on, presenting the choices in a manner where the ``optimal'' decision is made easier for users, and providing better default options to establish better privacy norms.

\subsection{Prior Work on Trackers}
Despite a wide range of news outlets reporting on different abuses of personal item tracking devices, there is very little prior academic work on this topic. Motherboard~\cite{motherboard22police} reported on 150 police records of AirTags being used in IPA. Apple was sued later in 2022 by two women who were stalked by their ex-partners, as reported by the New York Times~\cite{nyt22applesued}. The devices were also planted on cars to steal them later, as reported by Mac Rumors~\cite{macrumors21cartheft} and the New York Times~\cite{nyt21cartheft}.

Following these reports, several groups performed informal tests using the devices for stalking. Hill~\cite{nyt22airtags} used a range of tracking devices on her husband to follow his movements throughout the day, and he was unable to locate many of the planted trackers. Fowler~\cite{washingtonpost21frighteninglyeasy} consented to be tracked by a colleague to put the anti-stalking features to the test and found them severely lacking. Scott~\cite{tomscott22tracker1,tomscott22tracker2} ran a pair of games where one participant attempted to complete tasks while the other used an AirTag to try and track them down. In several cases, the tracking participant came very close to finding the other player.

Wired~\cite{wired21gifttostalkers} have reported specific failures of the Apple anti-stalking features that made them a ``gift to stalkers'', including quiet and infrequent sound notifications, long periods before alerting users, and the absence (at the time) of any Android anti-stalking features. Apple~\cite{apple22improvements} responded by creating the ``Tracker Detect'' app for Android, increasing the tracker volume, and reducing the period to notify users of unwanted trackers.

Heinrich et al.~\cite{heinrich22airguard} compare Apple's iOS tracker detection with their AirGuard app. By reverse engineering the functionality of Apple's iOS AirTag detection, they created AirGuard to scan for AirTags and later Tile trackers. Their application detects unwanted trackers significantly faster than official options.

Turk et al.~\cite{turk23cantkeepthemaway} analyse the implementations of anti-stalking features of item-tracking devices through a series of experiments with trackers combined with the authors' expert opinions on the topic. They identify a series of limitations in the current anti-stalking features, as well as limitations on their future design. They provide a series of potential improvements that could be made to anti-stalking features in the future. 

The limitations identified by Turk et al.~\cite{turk23cantkeepthemaway} include various issues around how devices scan for unwanted trackers. They find that the time taken to identify unwanted trackers with background scanning is too long, while manual scanning is too difficult to use effectively due to the extensive user effort required. There are also problems with the availability of anti-stalking features: as shown in Table~\ref{tab:tracker-comparison}, background scanning is not available for all operating systems and devices, and users are required to install multiple applications to be able to detect all unwanted tracker types. Additionally, they identify shortcomings in the methods to locate unwanted trackers. Using Bluetooth to locate trackers is not always provided as an option to users, nor is Apple's precise location feature. The sound alerts are also too easily muffled in common hiding spots used by stalkers. 

There are also fundamental limitations of the effectiveness of anti-stalking features. In addition to the malicious use for stalking, there is the legitimate use case of locating stolen items~\cite{turk23cantkeepthemaway}. This appears as though the tracker is stalking the thief and the anti-stalking features will help them remove the tracker. There is no way to distinguish this from malicious use, and companies are therefore forced to "pick a side" --- anti-stalking or anti-theft. Apple advertises that the AirTag is not intended to be used to locate stolen items, while Tile created an anti-theft mode which disables their anti-stalking features in favour of anti-theft. Avoiding other forms of false positive, such as using public transport with someone who has a tracker but whose Bluetooth is turned off, also limits the design of anti-stalking features.

\section{Item Tracking Devices and Their Features}\label{sec:trackers}
We selected 5 types of trackers to be used in our study. We first ran various tests with the trackers to understand how the anti-stalking features work. We placed all trackers on three users to measure time to alert for different phones/OSes. We also provide 8 of each type of tracker to participants to track other players with a total of 40 devices. The trackers we selected for evaluation are the Apple AirTag, the Tile Sticker, the Galaxy SmartTag, the Chipolo One Spot, and the ``Keyfinder''. 

AirTags are the most popular tracker and have gained media coverage for their use in stalking. Tile is the largest existing brand of item tracker before the AirTag, and provides several models: the Pro, Mate, Slim, and Sticker. As we are focusing on the use in stalking, we chose the sticker, which is the smallest tracker and also has a sticky surface to attach the trackers to other's possessions. The Galaxy SmartTag functions similarly to AirTags: the associated brand of phones all provide location updates for trackers, making them more effective at finding lost items. Chipolo trackers are another popular brand of trackers not associated with a phone brand, comparable to the Tile trackers. For the fifth tracker, we selected the most popular (by ratings) off-brand tracker from Amazon as a control, which is the Kuxian Keyfinder. This tracker does not provide any anti-stalking features.

\begin{table*}[t]
    \centering
    \caption{Summary of Anti-Stalking Feature Availability}
    \begin{tabular}{|c|c|C|C|C|C|C|}
        \hline
        \multicolumn{2}{|c|}{Application} & Apple: Find My & Apple: Tracker Detect & Samsung: Smart Things & Tile & AirGuard\\
        \hline
        \multirow{2}{*}{Available on} & iOS & Built-in & \no & \yes & \yes & \yes \\
            & Android & \no & \yes & \yes & \yes & \yes \\
        \hline
        \multirow{4}{*}{Detects}  & AirTag & \yes & \yes & \no & \no & \yes \\
            & SmartTag & \no & \no & \yes & \no & \no \\
            & Tile & \no & \no & \no & \yes & \yes \\
            & Chipolo & \yes & \yes & \no & \no & \yes \\
        \hline
      \multirow{2}{*}{Scan Type} & Background & iPhone Only & \no & Galaxy Only & \no & \yes \\
            & Manual & \no & \yes & \yes & \yes & \yes \\
        \hline
    \end{tabular}
    \Description{The set of features available for each of 5 anti-stalking applications. Apple provides Find My for iOS and Tracker Detect for Android. Both can detect Airtags and Chipolo trackers, but only the iPhone Find My can use background scanning. Samsung provides Smart Things for both OSes, but only detects SmartTags and only provides background scanning to Samsung Galaxy phones. Tile and Airguard work on both OSes. Tile provides manual scanning for Tile trackers only, while Airguard provides both background and manual scanning for Airtags, Tile trackers, and Chipolo trackers.}
    \label{tab:tracker-comparison}
\end{table*}

These item tracking devices provide several features to help the owner locate them, in addition to anti-stalking features that build off of these. The trackers provide updates on their location when in Bluetooth range of the owner and remotely through several methods, and allow the user to locate the device when in range through Bluetooth and by making the device play a sound. The anti-stalking features revolve around providing mechanisms to scan for unwanted tracking devices, and then allowing the user to use the aforementioned features to locate any detected devices. Additionally, Apple provide ``separated from owner'' sound alerts periodically when the tracker is unable to connect to the owner's device for an extended period.

\subsection{Location Updates}
All of the trackers in our study use Bluetooth to provide the location of the tracker; they do not have any on-board GPS or other mechanism to update the owner on their current location. 
When they are in Bluetooth range of the owner's device, they are able to connect to inform the owner that the tracker is near their current location. We measured the Bluetooth range of each tracker and found that most trackers have a range of 55-75m; Apple's AirTags are an outlier to this at 26m.

Most trackers provide a mechanism to remotely update the location of the tracker. When the tracker moves in range of a compatible device, the location of the tracker is encrypted and uploaded to the manufacturer's servers, who then forward it to the owner to decrypt. Apple Find My compatible devices (Airtags, Chipolos) will provide this update when in range of any Apple device which is part of the network\footnote{All Apple devices are part of the Find My network by default. Users are able to opt-out in their device's settings menu.}, while SmartTags can report via any Samsung Galaxy phone. Tile trackers provide updates through other Tile app users.

\subsection{Scanning for Trackers}
There are two types of scanning available for unwanted trackers: background scanning and manual scanning. The availability of these scans is shown in Table~\ref{tab:tracker-comparison}. Background scanning listens for nearby tracking devices and records when and where they have been detected near the user. When a tracker has been detected in multiple locations over an extended period and distance, a notification is shown to the user stating that an unknown tracking device is following them. Apple provides background scanning for Find My compatible devices for iPhones, and Samsung provide background scanning for Galaxy SmartTags for Galaxy phone users with the SmartThings app installed.

The other type of scan is manual scanning. This requires the user to initiate a scan through the provided application and then move around during the scan. The feature identifies which trackers, which do not belong to the user, are moving with them, and provides a report on the unwanted trackers. This requires multiple scans from different locations to avoid false positives and privacy invasions of other tracker users who are nearby when the user runs the scans. Apple provides manual scanning to Android users through the Tracker Detect app; Tile provides manual scanning for all users of the Tile app; and Samsung provides manual scanning for Galaxy users through the SmartThings app.

\subsubsection{Time to Alert}\label{subsec:times-to-find-trackers}
The time to inform users of unwanted trackers is not publically reported by any item tracking device manufacturer. Heinrich et al~\cite{heinrich22airguard} created test scenarios to measure the time it takes Apple to report an unwanted tracker and compared it with their own AirGuard app. They found that Apple's tracker scanning located the unwanted devices after 1 hour 45 minutes in the first scenario, 4 hours 14 minutes in the second scenario, and failed to detect them in the third scenario. Their own AirGuard app detected the trackers in 35 minutes, 30 minutes, and 4 hours 18 minutes respectively.

We created our own tests to measure the times taken to detect each type of tracker with background scanning. We provided an iPhone user, a Galaxy user, and a Google Pixel user with all five trackers and installed the SmartThings, Chipolo, Tile, and AirGuard applications. Two users are authors of this paper, while the third was selected due to the type of phone they had, and the range of activities they would be undertaking throughout the day. Each user then went about their regular activities and measured the time for each application to alert the user of each unwanted tracker, if at all. Note that the AirGuard application is not available on iPhones, and Find My scanning is not available on Android phones. These tests took place in February 2023. The reports received are in Appendix~\ref{app:reports}.

We found that very few of the trackers were detected, as shown in Table~\ref{tab:tracker-detection-times}. The iPhone alerted the owner to the unwanted AirTag after 5 hours 41 minutes, shortly after connecting to the home Wi-Fi, but failed to detect the Chipolo tracker. AirGuard detected the AirTag and Tile tracker simultaneously on the Pixel phone after 25 minutes, but missed the Chipolo tracker and did not find the tags on the Galaxy Phone. In all cases, the Samsung Smart Things application did not find the Galaxy SmartTag.

\begin{table*}[t]
    \centering
    \caption{Volume of Each Type of Tracker in Decibels. Background Volume is 34dB.}
    \begin{tabular}{|c|c|c|c|c|c|}
    \hline
        \multirow{2}{*}{Tracker} & Advertised & \multicolumn{4}{c|}{Maximum Measured Volume in (dB)} \\
        & Volume (dB) & Open Area & Coat Pocket & Bag (at 10cm) & Bag (at 5m) \\
        \hline
        AirTag & 60 & 86 & 76.3 & 40 & 36.5\\
        Galaxy SmartTag & 85-96 & 89 & 84 & 68 & 54\\
        Chipolo One & 120 & 96 & 80 & 64.5 & 47.9\\
        Tile Sticker & 85-114 & 92 & 76 & 61 & 40\\
        Keyfinder & --- & 85 & 70 & 55 & 38 \\
        \hline
    \end{tabular}
    \Description{The volume of each of the 5 trackers as advertised by manufacturers, and then as recorded in an open area, inside a coat pocket, inside a bag from 10centimeters away, and inside a bag from 5 meters away. All trackers have similar measured volumes, decreasing in each scenario; in the best case, all trackers are around 90dB (similar to a hairdryer), while in the worst case most trackers are around 40dB, barely above the background volume of 34dB.}
    \label{tab:tracker-volumes}
\end{table*}

\subsection{Finding Detected Trackers}
Nearby trackers, either the owner's or detected trackers following a user, can be located through two methods. Bluetooth provides the distance to a connected device, which users of any tracker in our study can use to get closer to the tracker once in Bluetooth range. Apple AirTags additionally allow the user to see the direction to a tracker through the Ultra-WideBand (UWB) support in the M1 chip, as long as the user's phone supports UWB (i.e. iPhone 11 and later).

The other primary method for locating nearby trackers are sound alerts. When a tracker is nearby, users can make it play a sound so that they can easily locate it. We measured the volume of the different trackers in several common scenarios, shown in Table~\ref{tab:tracker-volumes}. These include common hiding places such as coats and bags; we also measured the volume at varying distances in the latter case. While the trackers are relatively loud for their size, the sound alerts are easily muffled to be quieter than speech (approximately 60dB).

Another additional feature provided by Apple AirTags are separated from owner sound alerts. When a tracker has been separated from its owner for a random period between 8 and 24 hours, the AirTag will play a 15 second sound alert every 6 hours. It will additionally attempt to send a notification to nearby iPhones at the same time. These alerts are intended as a fallback in case scanning mechanisms fail, but the short sound and long delay between alerts severely limit the utility of this feature.

\section{Methods}
\label{methods}
Our study extends the game of ``Assassins'' by providing participants with trackers to place on other ``tracker-enabled'' players. Several incentives were provided to encourage use of trackers. Players are rewarded for completing our ``post-mortem'' survey, requesting information about how the tracker was used and how the tracked player (attempted to) locate it.

\subsection{Ethics}
\label{ethics}

We obtained ethics approval from the Cambridge University Department of Computer Science Research Ethics Committee before commencing our study. Our main ethical considerations included informed consent, possible misuse of trackers, recruitment and incentives, data collection and storage, and anonymisation.

All participants provided informed consent before any information was collected. The consent form explains the study, the data collected and how it is later anonymised, how to withdraw from the study, and who to contact with further questions. This immediately preceded the demographic data collection. Participants explicitly consented to be tracked and to only track other tracker-enabled players. They agreed not to use the tracker on people or items that were not theirs after the study. They also had to inform us of the tracker's location after it had been planted in case it needed to be retrieved.

Participants were recruited through the Assassins' Guild student society at the University of Cambridge. Students typically sign up for the game online or at the annual Freshers' Fair held at the start of the academic year. The study was advertised alongside the game, and all players who signed up for the game were able to sign up as tracker-enabled (provided they completed the consent form) or tracker-disabled. We allowed players who were assigned trackers to keep them after the game to incentivise people to join, in addition to providing bonus points for using trackers.

We collected participant contact information, demographic information, and prior experience with technology and the Assassins' Guild. We also collected participants' responses to the post-kill surveys throughout the term. Data were not anonymised during the course of the game as we needed to maintain contact to gather survey responses. At the end of the game, we pseudonymised the results by replacing player names with unique identifiers.

Participants are able to contact the researchers at any time to request a copy of their data or to withdraw from the study. We established a protocol for withdrawal if it were to occur while the game was underway: we would delete all data stored about the participant and contact the person tracking them to ensure that the tracker was returned. We would also obtain the current location of the participant's tracker so that we could remove it.

Despite the precautions taken, there was still a possibility that a participant could accidentally place a tracker on someone who had not consented to being tracked. We created labels for the trackers featuring the logo of the Assassins' Guild and a QR code to a webpage\footnote{\url{https://www.cl.cam.ac.uk/~kst36/tracker_study.html\#found-tracker}} 
which provides details on the study, contact details for the researchers, and instructions to disable an unwanted tracker.

\subsection{The Assassins Game}
In a game of Assassins, players are provided with three targets to track down and ``kill'', by using fake weapons such as a ``knife'' (pen or another object with ``knife'' written on it) or a water pistol. There are restrictions on when assassins can make kills, such as not playing the game in defined out-of-bounds areas like lecture theatres or in spaces where the general public may become concerned. 
After a kill, players write reports on both the accurate list of events and a dramatic retelling, which are published on the Assassins' website. In addition to having three targets, players are the target of three other players. The targets are replaced after each kill, and a new player will be assigned to track a target if one of the players hunting them dies.

Players are rewarded for making attempts and getting kills through ``competence''. When a player's competence runs out, they become ``incompetent'' and their information is published on the Assassins' website, making them a target for all players. Making two attempts or getting a kill adds a week to the player's competence period, incentivising players to remain active. In addition to regular players, there are ``Police'' who may only ``kill'' anyone on the incompetents list. Any members of the Police who break the rules will become wanted and are subject to similar rules to incompetent players.

The last week of the game is ``open season'', in which all remaining players' information is provided on the website and players can target any other player still alive. A scoreboard is also published, and players can attempt to get into the top 6 who will go on to duel to win the game.

In our study, players can sign up as ``tracker-enabled'' or ``tracker-disabled''. 40 tracker-enabled players were randomly selected to receive trackers which were compatible with their phone. To incentivise the use of trackers, players were provided with twice the competence bonus and four times as many points if they planted a tracker on the target and left it for at least 4 hours before the kill. Tracker-enabled players without trackers were invited to remain tracker-enabled and were rewarded with additional points for each kill they made.

Players are explicitly told which of their targets are tracker-enabled and tracker-disabled. Players on the incompetent and wanted lists are marked with this information. Tracker-enabled players therefore have one, multiple, or no players tracking them at any given time.

Players taking part in the study were intentionally not informed about the available anti-stalking features. This makes the study naturalistic and allows us to better understand how people will attempt to find trackers in the real world, rather than pushing the use of technical countermeasures.

\subsection{Trackers}

To distribute the 40 trackers, we randomly allocated each type of tracker amongst participants with compatible phones. This information had been collected from participants at the time they signed up to take part in the study. We first allocated 8 AirTags to randomly selected iPhone users, then 8 Galaxy SmartTags to Galaxy users, then assigned all remaining trackers amongst all remaining participants.

\subsection{Surveys}
There were three surveys in our study. The first was completed at the time of recruitment. A ``post-mortem'' survey gathered information during the game, after each kill. A post-game survey was completed once the game had finished for the term.

The initial survey was completed after obtaining informed consent from participants, who were provided with detailed information about the study and how the game works when they are part of it. The survey asks for information on the participant's devices and operating systems, prior experience with the Assassins' game and any prior use of trackers.

The ``post-mortem'' survey was 
completed after one player ``kills'' another during the course of the game while using a tracker. It asks the person using the tracker how they planted it and how easy and effective it was for stalking. Both the tracker and tracked player were asked to provide information on how they searched for trackers that had been planted on them. They are asked to provide the current location of their own tracker so that it can be returned to the owner.

The post-game survey asks a series of questions initially overlapping with the post-mortem survey. The intention of this part is to gather the same information on how trackers were (intended to be) used, as well as how players were checking for trackers, even if they did not get the opportunity to use a tracker and/or were never tracked by other players. Additionally, the survey asks participants if they used each anti-stalking service, and if so how useful was it. Prior to this, participants had not been explicitly told what anti-stalking features were available to them, so this allows us to measure how often each tracker's anti-stalking features are used.

\section{Results}
\label{results}

In this section we discuss the results of our study with respect to the usage of trackers within the game and the results of our surveys. \S\ref{subsec:demographics} describes the background of the participants and \S\ref{subsec:tracker-use-in-game} discusses how trackers were used in the game. \S\ref{subsec:tracker-use-stalking} explores how effectively players used the trackers to locate other participants. \S\ref{subsec:locating-trackers-in-game} discusses how participants attempted to locate trackers planted on their person, and the use of anti-stalking features by participants.

\subsection{Demographics}\label{subsec:demographics}
372 players signed up to the overall game. Of these, 91 signed up as tracker-enabled and completed the consent form, and we provided 40 of them with trackers. The remaining 51 all opted to remain tracker-enabled despite not having a tracker (i.e. consented to remain as possible targets for tracking), in return for bonus competency points for their own kills.

Of the 91 tracker-enabled players, 45 (49.4\%) had iPhones, 16 (17.6\%) had Samsung Galaxy phones, and 30 (33.0\%) had another brand of Android phone. 67\% of participants were playing for the first time, 29\% had played between 1 and 3 games before, and 4\% had played 4 or more games before.

Most participants in our study are studying STEM subjects. Participants rated themselves as having above-average knowledge of technology (mean +0.52: 1 far below average, 8 somewhat below average, 31 average, 45 somewhat above average, and 6 far above average). 
although most participants had limited experience with tracking devices (57 had no experience, 25 had a small amount of experience, 8 had a moderate amount and 1 had a great deal of experience with using tracking devices).

\subsection{During the Game}\label{subsec:tracker-use-in-game}
Trackers were assigned when signup closed and delivered to participants within 12 hours of the game starting. Participants were then notified if they were given trackers to ensure they could begin to use them as soon as possible. Participants were also periodically reminded throughout the study to use the trackers and fill in the post-mortem survey to boost participation and avoid participants forgetting about the tracker component of the game.

One player successfully used the trackers twice to track down targets before achieving in-game kills, and two other players managed to plant their trackers although they were unsuccessful in using them before being killed.

In the first successful use of a tracker, the participant planted the tracker on the target when walking past them:
\begin{quotation}
    \textit{I placed the tracker in the open side pocket of their bag whilst passing them in a public place as they were walking somewhere from their lecture.}

    \textit{I used the tracker to check which lecture theatre they would be in as I knew they must be in one of 2 [from the location provided by the app].}
\end{quotation}
This allowed the participant to accurately locate where the target was, and to wait outside for them to appear. Later in the game, the same player managed to plant their tracker on another target:
\begin{quotation}
    \textit{When I arrived at their room, the key was left in the door. I attached the tracker to their keychain to find them later.}

    \textit{I found out they were at a meeting, then found out they were riding a bike due to how fast they were. I headed to intercept them at the bike rack closest to their room.}
\end{quotation}
In this case, the participant determined the mode of transport being used from the location updates of the tracker. From this, they were able to predict where they could go to intercept the person, before using other features of the tracker to identify exactly who it is:
\begin{quotation}
    \textit{I decide to turn on the tracker's alarm. I go to stab the person I believed was [the target], but moments before, the alarm rings on a blonde-haired girl nearby, saving the innocent and myself.}
\end{quotation}
Despite not having seen the person they were targeting beforehand, the tracker's sound alert allowed the participant to identify who they were tracking and avoid assassinating the wrong person.

\subsection{Using Trackers for Stalking}\label{subsec:tracker-use-stalking}
Many of the participants in our study found it difficult to plant trackers on their targets. Of 9 participants with a tracker who responded to the end of study survey, 7 attempted to plant their tracker at least once, and three were successful. Participants explained that the difficulty of planting trackers stemmed from attempting to plant them while in public spaces --- it was too easy to be spotted by their target or by the general public while planting the tracker, and there was the possibility of either the in-game or real-world police spotting the tracker being planted.

One participant was additionally concerned by an auxilliary feature of the Chipolo tracker (also provided by Tile and the Keyfinder): when the tracker is clicked twice, the owner's phone will ring out, ignoring if the phone is on silent. The possibility of being discovered this way was deemed too high of a risk by the participant, who then did not plant their tracker. The participant did not discover that this can be disabled in the Chipolo app.

Several participants attempted to plant trackers but found it difficult and were unable to plant them successfully. One attempted to plant an AirTag on a bicycle but were unsure how to attach it. Two others had difficulty finding a situation where they would be able to avoid raising suspicion from the public while also avoiding interaction with their target.

\begin{quotation}
    \textit{It was difficult in many situations to do so without arousing suspicion from other people.}
    \\
    \textit{Planting trackers is also way too difficult, and I needed multiple attempts before I could actually find my target unaware and for me to actually feel like the police wouldn't be called if seen.}
\end{quotation}

There were some successful uses of trackers in our study. Two participants planted trackers in others' backpacks, one player attached it to the underside of a bike seat, and another attached it to their target's key chain. Two of these cases were with a Galaxy SmartTag, one was an AirTag, and one was a Chipolo One Spot. Two of these players were unable to use the trackers on their target before being killed: one player complained that the Chipolo tracker was not providing any remote location updates:
\begin{quotation}
    \textit{The tracker was absolutely useless. As soon as the target went further away from my Bluetooth range, the app had no idea where the tracker is.}
\end{quotation}
The participant was unable to locate the tracked player before being attacked by in-game police. The other was attacked during the 4 hour waiting period after planting the tracker. The locations used to plant trackers are largely a reflection of the participants of our study being students, who are in constant possession of backpacks and commonly use bikes to travel.

The most effective trackers for stalking were the AirTag and Galaxy SmartTags. Both of these trackers are tied to a common phone brand (half of the participants had an iPhone, while 17.8\% had a Galaxy phone). This made the trackers provide more frequent updates, in one case often enough to allow the owner to determine what mode of transport the tracked target was using and predict where they were heading to intercept them.

\subsection{Locating Planted Trackers}\label{subsec:locating-trackers-in-game}
Most players in our study either did not search for trackers planted on them or relied on periodic manual searches of their possessions. Of 19 respondents to the post-game survey, 9 reported checking inside their bags, coat pockets, under their bike seats, and otherwise manually looking for trackers. Only one player out of 91 participants attempted to use any of the anti-stalking features provided by manufacturers, although several noticed that their iPhones automatically scan for AirTags. None of the players reported that any anti-stalking feature alerted them to the presence of an unwanted tracker.

Several participants did not search for trackers that may have been planted on them, even though they knew that they could be tracked at any time and had an incentive to prevent tracking. Commonly reported reasons for this were that they had realised how hard it was to plant trackers and had little concern about trackers being planted on them, or that they felt manual searching was sufficient. The participant who used some anti-stalking features discovered Apple's Tracker Detect app as well as the third-party AirGuard app and used the background and manual scans available in the respective applications, although they did not know about the Tile, Chipolo, or Galaxy anti-stalking features. Although they used the two applications to search for unwanted trackers, they did not locate any trackers planted on their person. This participant rated themselves as having ``somewhat above average'' technical knowledge, although they had never used a tracker or participated in an Assassins' game before this study. This could suggest that only the more technical users will use these applications, although many other participants in our study made use of them.

Some participants were aware of anti-stalking features but did not use them. Of 19 respondents to our post-game survey, 4 knew of at least 1 anti-stalking feature but none of them used the features. None of the respondents knew about the Samsung SmartThings application.

The most common anti-stalking feature used was the iPhone background scanning for Airtags. Even though participants did not look for technical means to detect trackers planted on them, Apple has unwanted AirTag detection built in to the OS of their phones, which means that users of these devices already had some anti-stalking protection by default, requiring no manual intervention.

\section{Discussion}
\label{discussion}

There are a number of implications of the results of our study. In this section, we discuss the scenarios in which trackers are more easily abused for stalking, in addition to which devices are easiest to abuse and why. We then describe the key limitation of anti-stalking features identified in this study, as well as how default options impact this. We finish by discussing the industry response to our work.

\subsection{Misuse Scenarios}
Many participants in our study reported that it was difficult to plant trackers due to having to attempt to plant them in public spaces. This is analogous to how stalkers may attempt to use trackers, and reflects issues they would also face: being spotted while planting a tracker, or being unable to access the victim's possessions to place a tracker as the victim is in close proximity to them. The difficulty may vary depending on if the stalker is targeting a specific person, or opportunistically stalking a victim when presented with a moment to plant the tracker surreptitiously.

On the other hand, it is much easier to plant these trackers in domestic abuse scenarios, where the abuser has access to all of the victim's possessions and can easily plant a tracker without the victim knowing. This is reflected by real-word reports of tracking devices being used for stalking: the vast majority of stories of their misuse are in domestic abuse cases, while few cases see trackers used on strangers.

\subsection{Ease of Abusing Trackers}
The effectiveness of item finding devices for stalking depends primarily on the remote location update mechanism used. Trackers which report their location when near to a phone of the same brand (AirTags and Galaxy SmartTags) are most effective for tracking, due to the frequent updates of the device's location. Other trackers are less effective as it is rare for them to update, which made them useless for participants in our study.

While improved remote tracker location enables malicious use, it also has legitimate benefits. If a personal possession is left at an unknown location, the owner can see where it has been left; if it is lost by airport or mail staff, the owner can inform the company of the current location; and if an item is stolen, it can be tracked down through remote tracking features.

\subsection{Almost No-one uses Anti-Stalking --- Even When They Know They're Being Stalked}
Our most striking finding is that of the 91 participants who knew they could be stalked as part of this study, only one participant found available anti-stalking features and attempted to use them. Other participants either searched their possessions manually or did not search for trackers at all. In a real-world scenario, people will not know if they are being tracked, however in our study all participants consented to be tracked beforehand and were explicitly made aware that they were likely to be tracked during the study. While iPhone users have background scanning by default, this only detects trackers that use the Find My network. In our evaluation, the participants were incentivised to detect and remove any trackers. However, they are rarely aware of or use the anti-stalking features provided by manufacturers --- which means that in the real world, even fewer people are likely to use anti-stalking features.

It is worth noting that in our study, no participant reported using the anti-stalking features successfully to locate a tracker. This is primarily due to the lack of successfully planted trackers in combination with very few players using anti-stalking features. This limits the evaluation of anti-stalking features to their availability rather than effectiveness in our experiment.

\subsection{Defaults}
One limitation of the Samsung SmartThings app is that background scanning is disabled by default. While it is great that background scanning is available, the easier option presented to users is to manually scan for trackers; they have to go into settings to enable background scanning for unwanted trackers. In contrast, background scanning for AirTags is provided by default for all iPhone users, even if they are not aware of anti-stalking features.

\subsection{Limitations and Strengths}
This study has attempted to overcome the significant difficulties associated with this challenging area of research. However, a number of limitations remain. Despite having 91 tracker-enabled players and 40 trackers, we only saw 4 trackers successfully planted on targets during our study. While this limits the conclusions that can be drawn about the effectiveness of each anti-stalking feature, we note the purpose of the study was to explore users' experience with anti-stalking features, rather than evaluate the usefulness of the trackers themselves.

Our study does, however, have a number of strengths. We designed the research with ethical considerations at the forefront, with participants consenting to being tracked. The study was naturalistic, in that participants did not know when they were being tracked (the study ran for 8 weeks), who by, or by what type of tracker. This is consistent with how trackers may be used for stalking. The study was also gamified, with participants being rewarded for using and detecting trackers, which strengthens our findings about their lack of uptake of anti-stalking features. 

In our study, participants were explicitly made aware that they would be tracked by other players to ensure that they can properly consent to this study. Although this makes the study slightly less naturalistic, in that people in the real world will not know if and when they are being tracked, this does imply that even fewer people in the real world will use the anti-stalking features.

The participants in our study are university students primarily from a STEM background with a self-reported higher than average technical knowledge. This implies that the participants are more likely to be able to use the trackers, and anti-stalking features, more effectively than the average user. This suggests that in the real world, even fewer people would make use of the anti-stalking features than in our study --- however, almost no participants made use of them. The demographic of our participants, combined with them explicitly being made aware of possible tracking and not using anti-stalking in spite of this, suggests that the anti-stalking features are extremely unlikely to be discovered and effectively used by real world users.

\subsection{Reporting and Response}
The results and recommendations from this paper, alongside the results of Turk et al.~\cite{turk23cantkeepthemaway}, were communicated to the four companies which provide anti-stalking features in early March 2023. Apple responded with a list of improvements they made in June 2021 and February 2022 but did not directly respond to our results and suggestions. We additionally reached out to Google with our results, aiming to integrate anti-stalking into the Android OS to allow for universal anti-stalking features. We met with several people on Google's safety teams to discuss our work and recommendations, in addition to possible issues that may arise (such as false positives) and how to design around them.

In May 2023, Apple and Google announced a joint draft RFC~\cite{detecting-unwanted-location-trackers-00} to standardise anti-stalking features and how they are implemented. This includes standardising the background scanning in operating systems, ensuring all trackers provide methods to trigger sound alerts, suggesting other mechanisms for locating identified unwanted trackers in future designs, and ensuring trackers provide a feature similar to Apple's away-from-owner alerts. The standard includes all of our suggestions for improving existing implementations. Other companies including Samsung, Chipolo, Tile, Pebblebee and eufy Security have expressed interest in this draft, suggesting they will incorporate the changes once the standard is complete.

\section{Conclusions}
\label{conclusion}

In this study we have analysed the use of personal item trackers for stalking and analysed the use of anti-stalking features. We see that the devices are more easily used in intimate partner abuse than for stalking strangers, although both are risks. We found that even in the ideal case of technically skilled people knowing that they will be stalked and what by, very few people use or are even aware of anti-stalking features available to them, implying that the feature sees little to no use in the real world. This, combined with other failures of the anti-stalking features, implies that they are little more than security theatre and are intended to quell concerns rather than to improve user safety. We provide recommendations for improving the availability of these features so that they are more accessible to the users who need them.

\begin{acks}
We thank our colleagues at the Cambridge Cybercrime Center for their feedback, and the Cambridge University Assassins' Guild for collaborating on this project. We would also like to thank Henry Caushi for inspiring the project.

This work was supported by the UK Engineering and Physical Sciences Research Council (EPSRC) grant EP/T517847/1 and the European Research Council (ERC) under the European Union’s Horizon 2020 research and innovation programme (grant agreement No 949127).
\end{acks}

\raggedright
\bibliographystyle{acm}
\bibliography{trackers}

\appendix
\section{Tests Scanning for Known Trackers}\label{app:reports}
\begin{table*}[h]
    \centering
    \caption{Detection Time for each Tracker by each Device using Background Scanning}
    \begin{tabular}{|c|c|c|c|c|c|c|}
        \hline
        Tracker & \multicolumn{2}{c|}{AirTag} & Tile & Galaxy & \multicolumn{2}{c|}{Chipolo}\\
        \hline
        Detected By & Find My & AirGuard & AirGuard & SmartThings & Find My & AirGuard \\
        \hline
        iPhone & 5h 41m & --- & --- & Not Found & Not Found & ---\\
        Galaxy & --- & Not Found & Not Found & Not Found & --- & Not Found \\
        Pixel  & --- & 25m & 25m & Not Found & --- & Not Found \\
        \hline
    \end{tabular}
    \label{tab:tracker-detection-times}
\end{table*}
\subsection{User 1: Galaxy Phone}
\begin{quotation}    
\textit{I have a Samsung Galaxy A40. My phone’s Bluetooth and location were switched on all day. 
The trackers were planted at 9:45am, in the department next door to mine. I then walked to my department, and the trackers remained in my bag all morning, with my phone less than a metre away most of the time (the only exceptions being three trips to get coffee from the departmental machine taking a few minutes each).
The phone automatically connected to the local University Eduroam WiFi and remained connected from 9:45-1pm.
At 1pm, I took my bag (and trackers) to the gym to get changed for a run. The walk to the gym took 5 minutes with my phone’s network data being switched on. At the gym, the phone automatically connected to the local Eduroam wifi.
At 1:15pm I took the trackers with me on the entire 10.71 mile run (which took 1.5 hours). The trackers were stored in a pocket a few centimetres away from my phone at all times. My phone was connected to my phone’s network data at all times during the run (and bluetooth \& location were switched on). I didn’t stop at all during the 1.5 hours and ran across fields and through local villages during the run. I returned to the gym at 2:45pm at my phone automatically connected to the university Eduroam wifi. At 3pm I walked back to my University department and my phone again connected to the university Eduroam wifi automatically at 3:10pm. The trackers then stayed within 1 metre of my phone all afternoon (with the exception of two trips of less than 5 minutes duration each for coffee from the local departmental machine).
At 5.10pm I walked home. My phone automatically disconnected from university Eduroam wifi and my phone’s network data automatically took over. I arrived home at 17:39pm and my phone automatically connected to my home wifi. My phone remained connected to the home wifi all evening.
I didn’t receive any notifications regarding the tags at all. The Chipolo app briefly gave one notification at around 1pm that the Chipolo app was "running in the background" (at approximately the time my phone would have disconnected from Eduroam and connected to my phone’s network data when I started walking to the gym) but no notifications regarding tags detected. None of the other apps gave any notifications.}
\end{quotation}

\subsection{User 2: iPhone}
\begin{quotation}
    \textit{I have an iPhone X. I carried the trackers in my pocket all day. I cycled to the end of the street (so as to no longer be on my home Wi-Fi) before turning my Bluetooth on at 10:15am. I then continued cycling to my destination (2.1 miles). I was at this destination for around 40 minutes, before cycling to my next destination (2.3 miles). I left after approximately one hour and wandered around town on foot. I then cycled 1.3 miles to the grocery store and did a weekly shop, before cycling .3 miles home. 
    \\
    I returned home at 3:50pm. At 3:56pm I received a notification on my iPhone that it had detected an AirTag moving with me. I tapped on the notification, and it opened the Find My app. It first showed me a ‘what’s new in Find Me’ page. After I hit continue it showed me a map of my travels, and informed me it first detected the AirTag at 10:20am. One option presented was to learn more about the AirTag. I selected this and was instructed to bring the phone near the device, then tap on the on-screen notification. I tried to do this for several minutes, but no notification was delivered. 
    \\
    None of the other apps I installed detected any of the trackers. My iPhone only detected the AirTag, not the other trackers. It only notified me of this over 5.5 hours after I started being tracked, soon after I returned home and connected to the home Wi-Fi. }
\end{quotation}

\subsection{User 3: Google Pixel}
\begin{quotation}
    \textit{I have a Google Pixel 4. I obtained the trackers at my department and put them in the lower pocket of my trousers, below the pocket containing my phone. I kept my Bluetooth off until 4:01pm when I left the department. I then walked 1.4 miles to the city centre and began grocery shopping, and shortly after arriving I received 2 simultaneous notifications from the AirGuard app at 4:26pm. The notifications alerted me to an unknown AirTag and Tile Sticker respectively, and opening the notifications showed the path I had taken into the city centre. I then continued shopping around the city centre, walking approximately 1.6 miles total, before walking 0.8 miles home. I stayed at home for 2 hours, then walked the 0.8 miles back to the city centre for a social event. After several hours, I walked back home. Aside from the two early notifications, I was not alerted to any unwanted trackers present on me.}
\end{quotation}

\end{document}